# First detection of optical light from SNR G279.0+1.1

M. Stupar,[1,2] & Q.A. Parker[1,2]

[1] Department of Physics, Macquarie University, Sydney 2109, Australia

[2] Anglo-Australian Observatory, P.O. Box 296, Epping, NSW 1710, Australia

This is the initial paper in a series presenting the first optical detections and subsequent follow-up spectroscopy of *known* Southern Galactic supernova remnants (SNRs) previously discovered in the radio. These new detections come from the AAO/UKST H$\alpha$ survey of the Southern Galactic plane which has opened up fresh opportunities to study Galactic remnants. Here we present the first optical imaging and follow-up spectra of Galactic SNR G279.0+1.1 where a series of 14 small-scale fragmented groups of H$\alpha$ filaments have been discovered in a ~2.3° area centred on G279.0+1.1. Individually they are somewhat inconspicuous but collectively they are completely enclosed within the overall radio contours of this known SNR. Three of these filamentary groupings are particularly prominent and optical spectra have been obtained across two of them. Their morphological structure and spectral characteristics are typical of optically detected SNR filaments. Very strong [S II] emission relative to H$\alpha$ has been detected with [S II] / H$\alpha$ > 0.7 and 1.1, confirming strong, shock heated emission. This is sufficient to classify these filaments in the likely SNR domain and therefore indicating a direct connection with the radio remnant. Other typical SNR emission lines such as [O II] at 3727Å, H$\beta$, [O III] at 4959 and 5007Å, H$\alpha$ and [N II] at 6548 and 6584Å were also detected, lending strong support to an SNR origin of these optical filaments. The value and insights that these optical data can provide for known remnants are discussed along with their relevance to the Galactic nitrogen abundance. A serendipitous discovery of an adjacent H II region is also briefly described.

## 1 Introduction

Supernova are one of the most energetic processes in the universe and their subsequent remnants (SNRs) have a significant influence on the distribution of heavy elements and chemical enrichment throughout the Galaxy. Remnants are only visible for a few tens of thousands of years before they dissipate completely. This makes them a relatively rare Galactic population though they leave behind pulsars, neutron stars and black-holes as a legacy of their passing. Unsurprisingly, the highest concentration of supernova remnants (SNRs) in our Galaxy is within the Galactic plane where the massive progenitor stars of type II supernova at least are born and die before they can percolate out of the disk. Although they are extended objects, especially in their evolved state, their detection in previous extant broad-band optical wide-field surveys has been largely blocked due to the significant concentrations of absorbing gas and dust within the plane. However, SNRs are strong non-thermal sources and emitters of synchrotron radiation such that the vast majority of known remnants have been discovered and are best seen in the radio part of electromagnetic spectrum. They are detected from low to high frequencies where their measured negative spectral indices act as a key identifier. Consequently, they have formed a natural focus for radio surveys and follow-up radiO Investigations.

Green's on-line SNR catalogue (Green 2006) of $\sim 265$ carefully evaluated Galactic SNRs represents the most comprehensive compilation currently available. However, only $\sim 17\%$ of these catalogued Galactic SNRs so far have an associated optical detection. As this is the first in a series of papers dedicated to the discovery of optical counterparts to previously known SNRs a brief discussion of the importance of such optical data is given below.

## 2  Optical detection of SNRs

The optical study of Galactic SNRs permits a deeper understanding of the physical conditions in the remnant, such as density, kinematics and chemistry that is not possible with radio data alone. The optical data, when combined with radio, X-ray and infrared data, are of particular value as the relationship between shock excited gaseous emission and synchrotron radio emission can be investigated (Cram, Green & Bock 1998). With SNR optical identification we see a correlation between the shock wave and surrounding ISM, especially if the supernova blast propagates into dense, cool regions where H$\alpha$ emission is dominant. Optical SNR observations, especially detailed spectroscopy, allow studies of the change of chemical composition, density, temperature and their inter-relation as a function of SNR evolutionary state and environment. In the few Galactic SNRs with Balmer-line dominated filaments, such as seen in Kepler's remnant or Cygnus loop, the narrow band components of these lines from the (cooled) shock front can be compared with broader components produced in the hot area behind the shock which provides the best estimate of shock speeds in SNRs (Ghavamian et al. 2000).

Furthermore, though studies of Galactic abundance gradients are mostly based on giant stars or data from H II regions and planetary nebulae (PNe), SNRs have also been used for these estimates (e.g. see Gonzalez 1983) and despite the large data scatter can provide an additional useful constraint. The scatter has a strong statistical contribution due to the relatively small number of optically detected and studied SNRs currently available until now.

With optical SNR investigation we are typically refering to those which are in the mid to late phase of their evolution, such as the Vela SNR. Such SNRs represent more than 95% of the known Galactic population. The optical detection of such senile remnants is generally hard compared to the small number of young, optically bright Galactic SNRs such as the Crab nebula and the 300 yr old remnant Cassiopeia A, where both discovery and first observations were in the optical. However, even young remnants can be hard to detect optically, depending on Galactic sight line, proximity and environment. For example SNR Vela Z (G266.2-1.2, RXJ0852.0-4622) or Vela Junior, has an estimated age of only ~680 years (Iyudin 1998) but is not easily seen as a discrete entity in the optical though it is readily observed at a range of radio frequencies (e.g. see Stupar et al. 2005). This is because, unlike the coherent radio structure of Vela Z, the optical H$\alpha$ counterpart is in the form of fragmented filaments making it hard to disentangle from the highly confusing environment of the very extended main Vela SNR, perhaps the most famous optical SNR of all.

## 2.1  Searches for new SNR via their optical emission

Apart from the small sample of well know SNRs with optical counterparts, an active Greek group (e.g. Boumis et al. 2005; Mavromatakis et al. 2004 and references therein) have been undertaking targeted deep, narrow-band optical imagery in the region of known northern radio remnants and have been successful in uncovering new, optical SNR counterparts which they have followed up with matching optical spectroscopy. Optical emission in some remnants was also detected on the basis of their X-ray data (e.g. Gerardy & Fesen 2007). One can surmise that it is the local physical nature of the surrounding ISM, such as temperature, density and strength of any magnetic fields, that influences the correlation, or lack of it, between the detection of the non-thermal remnants in the radio continuum, and their optical counterparts in H$\alpha$ light (e.g. Cram, Green & Bock 1998). In many cases, the results seen in H$\alpha$ are fragmented components of the remnant in the form of filaments or more diffuse emission clouds (see example in Stupar et al. 2007b).

With the availability of the AAO/UKST H$\alpha$ survey of the Southern Galactic Plane (Parker et al. 2005), an extensive search of the resultant digitally rendered data in the form of the SuperCOSMOS H$\alpha$ survey (SHS) was undertaken for filamentary nebulosities as part of an on-going programme to identify new Galactic SNRs in the South. The SHS is currently the most powerful southern H$\alpha$ survey in terms of its combination of 5 Rayleigh sensitivity, 1 arcsecond resolution and 4000 sq.degree coverage of the Southern Galactic plane. Hence, its ability to reveal undetected optical counterparts of previously unrecognised SNRs is not altogether surprising. This has led to the initial identification of 18 new Galactic SNRs (Stupar, Parker & Filipović 2008) not previously identified in the radio or any other regime. The remarkable aspect of these discoveries, which have been largely confirmed by follow-up optical spectroscopy and post discovery radio recognition, is the fact that they were uncovered first purely on the basis of their optical images in the SHS data which is conveniently available online.[1] Many candidate remnants were uncovered in this way, supplemented by additional inspection of the original, archived survey film exposures at the Royal Observatory Edinburgh by MS

---

[1] http://www-wfau.roe.ac.uk/sss/halpha/

(refer to Stupar, Parker & Filipović 2007a for full details).

## 2.2 Searches for new optical counterparts to known radio remnants

During the extensive search programme describe above, targetted examination of the survey areas at the locations of known radio detected SNRs was also performed. It was during this process that many exquisite and large scale optical detections of known radio remnants were made for the first time, (e.g. refer to preliminary findings reported in the AAO Newsletter: Stupar, Parker & Filipović 2007c) where the first optical detections of SNRs G4.2-3.5, G15.1-1.6 and G315.1+2.7 are presented. The discovery of optical counterparts to other known radio remnants will be the subject of further papers in this series. Here, by means of introduction to these new detections, we concentrate on the known, highly polarised remnant G279.0+1.1. Note that the SHASSA H$\alpha$ survey of Gaustad et al. (2001), although of high sensitivity, has only 48 arcsecond resolution making it largely insensitive to the fine filamentary and fragmented structures of the newly discovered optical counterparts to known and newly detected SNRs uncovered by the SHS.

# 3 Optical detection of SNR G279.0+1.1

G279.0+1.1 was first discovered by Woermann & Jonas (1988) in radio maps at 2.3 and 1.6 GHz as an extended non-thermal source approximately 1.6° in apparent diameter. G279.0+1.1 is an unusual radio remnant due to a high level of polarization, up to 50% (Duncan et al. 19995), while SNR polarizations of <10% are more typical.

We present the first optical images of groups of H$\alpha$ filaments uncovered from the SHS in the vicinity of G279.0+1.1 and compare their spatial distribution with the existing radio data. This is followed with preliminary medium resolution spectra taken across two of the most prominent H$\alpha$ filaments which confirm shock excitation typical of SNRs. The derived physical properties of G279.0+1.1 are discussed together with brief mention of an unrelated optically detected H II region serendipitously discovered in the vicinity of the SNR and immediately to the west of the compact, strong extragalactic radio source G278.0+0.8 (Duncan et al. 19995) (see later discussion).

Identification of the three most prominent optical emission regions corresponding to the location of the radio remnant G279.+1.1 was made from the factor 16× blocked-down SHS survey digital data of field HA275. An association with the radio remnant was considered possible due to the large size of G279.0+1.1. Consequently, direct visual inspection at the ROE plate Library of the original survey films was undertaken to see if further optical filaments were present in the area of this SNR. This uncovered 11 more emission clouds and filaments (see Table 1) in the region. They vary in size, from between 1 and 5 arcmin and were easily recognized on the original films. Later their full resolution digital SHS counterparts were extracted for confirmation and quantitative evaluation of size and position. The overall morphological structure of these 14 groups of filaments/emission clouds in part follows the SNRs radio contours and are typical of optically detected SNRs. Hence, a probable association with G279.0+1.1 was assumed.

Fig. 1 shows the position of all 14 filaments against the radio contour image of G279.0+1.1 taken from the PMN 4.85 GHz radio survey (Condon, Griffith & Wright 1993). The three brightest groups of filaments are marked as F1, F2 and F3 on this figure. The positions of the other 11 nebulosities and filament groups are marked with small diamonds. It is clear that these probable optical counterparts are encapsulated within the overall ring shape of the fragmented radio contours of this remnant though clearly concentrated to the middle and North-Eastern edge of the radio shell. The reason for this becomes clear when a combined IRAS and radio colour composite image is created (see Fig. 2). The prevalence of dust in and around the southern and western rim of the SNR precludes easy optical detection. It is also interesting that the dust appears to 'respect' the overall SNR morphology on 3 sides indicating possible sweeping up of the ISM by the SNR along the S-W region in particular.

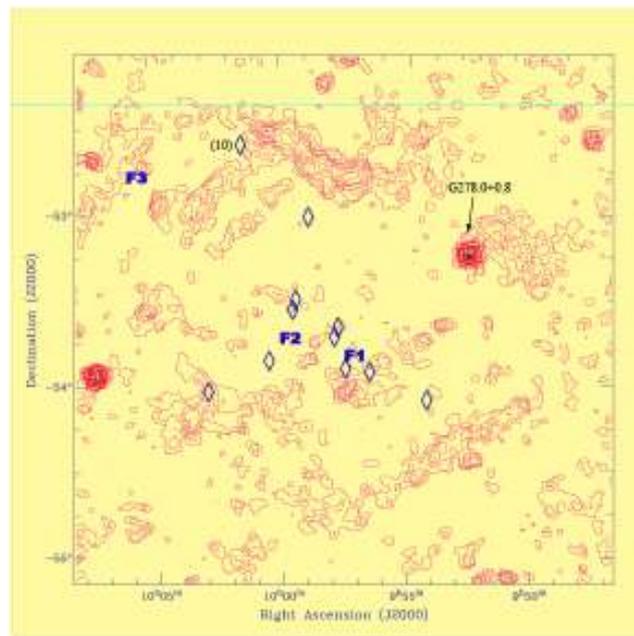

**Figure 1**: Radio contour image (contours from 0.01 to 0.30 Jy beam$^{-1}$) overlaid over the PMN (4.85 GHz) picture of G279.0+1.1. The brightest three filament groups used for further investigation are marked with (**F**). The other 11 groups mostly have very low surface brightness and are marked with a ♦ on the figure. An arrow marks the position of the G278.0+0.8 and separate H II region (see later discussion). Number (10) near ♦ sign in the upper left corner shows separate bright filament marked under number 10 in the Table 1 and also clearly seen on Fig. 3. The angular diameter measured from the top to the bottom of the remnant in a N-S direction is 2.3°.

The identification of 14 faint emission clouds and fine scale filaments again confirms the superior quality and sensitivity of the SHS since Woermann & Jonas (1988) specifically mentioned that they could not find any optical counterpart in their examination of extant emission-line atlases. Here the brightest three filament groups are examined more closely: two in the central area of G279.0+1.1 and the third at a distance of some 1.3° from this group on the N-E border of the remnant (see Fig. 1).

**Table 1**: Position of 11 low surface brightness clouds and H$\alpha$ filaments and approximate position of the filament group F2.

| Number | R.A. | Dec | Approx. size (arcmin) | Description |
|---|---|---|---|---|
| | h m s | ° ′ ″ | | |
| 1 | 09 54 04 | --54 04 04 | 1 | filament |
| 2 | 09 56 30 | --53 54 48 | 4 | filament |
| 3 | 09 57 19 | --53 53 08 | 5 | filament |
| 4 | 09 57 52 | --53 40 22 | 2 | cloud |
| 5 | 09 57 52 | --53 40 31 | 2 | filament |
| 6 | 09 59 02 | --52 58 28 | 5 | filament |
| 7 | 09 59 24 | --53 28 49 | 1 | filament |
| 8 | 09 59 40 | --53 33 43 | 3 | cloud |
| 9 | 10 00 37 | --53 52 48 | 2.5 | cloud |
| 10 | 10 01 54 | --52 36 09 | 2 | filament |
| 11 | 10 03 50 | --54 00 37 | 2 | filament |
| F2[*] | 09 59 45 | --53 45 00 | -- | see text |

[*]For slit position of filament groups F1 and F3 see Table 2

We also conclude from Fig. 1 and from inspection of the same area of the higher quality SUMSS radio survey[2] at 843 MHz (Cram, Green & Bock 1998) that the diameter of this remnant at these two frequencies is in fact 2.3°, somewhat larger than the originally determined size of 1.6° given by Woermann & Jonas (1988) (1.6 and 2.3 GHz) and Duncan et al. (1995) (1.4 and 2.4 GHz].

The field around filament group F1 (from Fig. 1) represents three filaments (lower right panel on Fig. 3). Two of the filaments in this group are ~13 and 3 arcmin in length respectively and are aligned in a N-S direction. There is one bright optical filament some 1 arcmin in length located at R.A.= 09$^h$ 57$^m$ 08$^s$ and $\delta$ = -54°02′57″ (with light extension of 3 arcmin in the west direction) which exhibits a classical SNR morphological structure. This panel also shows the spectrographs slit position across the chosen filament (see later discussion).

The region of filament group F2 (lower mid panel on Fig. 3) has a few short but prominent elongated nebulous structures, typical of optical SNRs, which are some 8 to 10 arcmin in extent.

There is a double arc structure with an additional filament distributed in the N-S direction some 13 arcmin in length which comprises filament group F3 (lower left panel on Fig. 3). This filament group is located on the N-E radio limb of G279.0+1.1 which also motivated us to take optical spectra there and to confirm the SNR nature of these outer optical filaments (see next section).

---

[2]http://www.physics.usyd.edu.au/ioa/Main/SUMSS

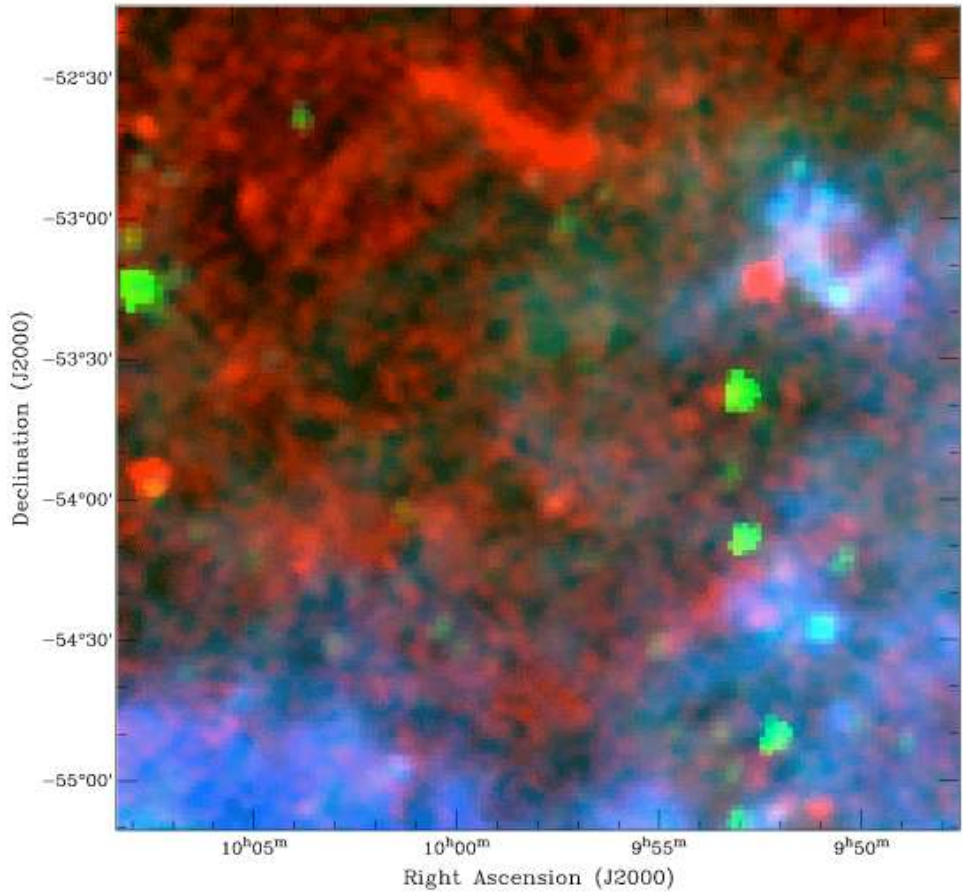

**Figure 2**: Combined IRAS 100 $\mu$m (light blue), 25 $\mu$m (green) and 4.85 GHz (red) radio image of the SNR. Note how the 100 $\mu$m IRAS dust signature encircles the lower SW side of the SNR and possibly explaining the lack of any optical SNR emission from this region.

### 3.1 Spectral observations

Spectral observations of G279.0+1.1 were performed on June 6, 2005 (see Table 2) using the Double Beam Spectrograph (DBS)[3] of the MSSSO 2.3m telescope. The DBS enables both red and blue spectra to be obtained simultaneously at different resolutions via a dichroic which splits visible light (between 3200 and 9000Å) to feed two essentially identical spectrographs. In the blue arm we used a 600 lines/mm grating which covered wavelengths from 3700 to 5500Å. In the red arm we used a higher resolution 1200 lines/mm grating with a spectral range between 6100 and 6800Å, sufficient to cover the key diagnostic emission lines and provide useful kinematical information. The slit width was set to 2.5" and resolutions of 2 and 1Å were achieved for the blue and red regions respectively. Standard IRAF slit-spectra reduction techniques were applied and the extracted spectra were flux calibrated via observations of the photometric standard star LTT7379. Although the observing night was not quite photometric some moderate extinction was estimated and the fluxes were corrected for interstellar extinction (see Table 3).

---

[3] http://www.mso.anu.edu.au/observing/2.3m/DBS/

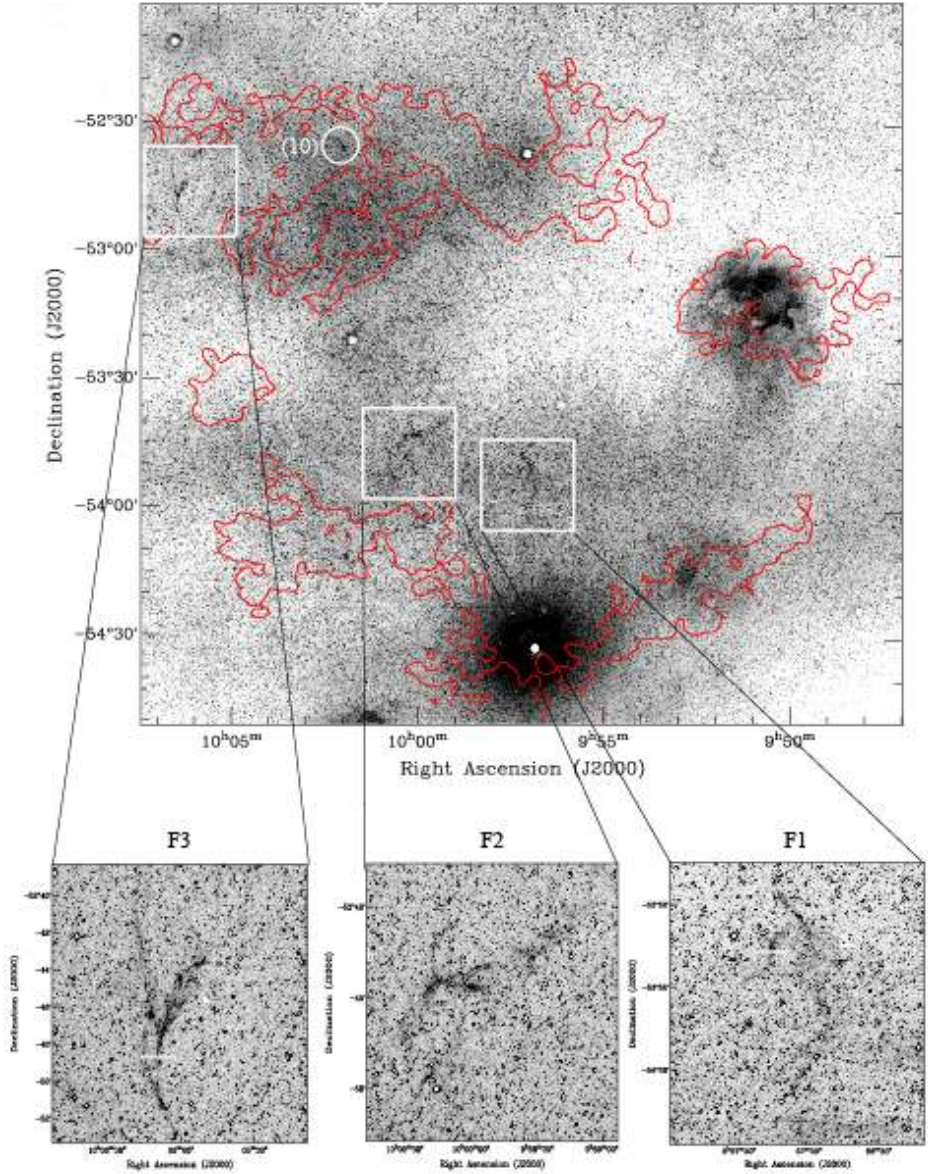

**Figure 3**: The upper 2.5° × 2.5° image is a low resolution quotient image of the SHS H$\alpha$ and SR image where all three filament groups are clearly seen, including a new H II region on the western region of SNR G279.0+1.1. For better definition, the lower figures represent enlarged images of F1, F2 and F3 filament groups also as quotient image but in high resolution. The strongest PMN radio contours from Fig. 1 are also included just to show the radio borders of G279.0+1.1. Radio features from inside area (and out of the SNR border) are removed to gain better definition of optical filament groups. On the large quotient image one can also notice strong separate optical filament (inside white circle), some ~2 arcmin in size which correspond to filament position 10 from Table 1 (see also Fig. 1). White rectangles on lower images represent position of the slits where spectra are taken for F1 and F3 filament groups (see also Table 2).

Errors were estimated for both the wavelength calibration and flux estimate. A mean rms of 0.03Å was obtained for the 600 lines/mm grating in the blue and 0.008Å for the 1200 lines/mm grating in the red. For the relative percentage error in flux estimate the brightest lines were employed which, for the 600 lines/mm grating, gave an estimate of ~15% error, and 18% for the 1200 lines/mm grating in the red.

**Table 2**: Spectral observing parameters for Galactic SNR G279.0+1.1. The 2.3m double beam spectrograph (DBS) was used for observations of F1 and F3 filament groups.

| Date observation | Grating (lines/mm) | Exposure time (sec) | Slit R.A. $h\ m\ s$ | Slit Dec. $°\ '\ ''$ |
|---|---|---|---|---|
| F1 06/06/2005 | 600 | 1200 | 09 57 16 | --53 53 02 |
| 06/06/2005 | 1200 | 1200 | 09 57 16 | --53 53 02 |
| F3 06/06/2005 | 600 | 1200 | 10 06 07 | --52 47 50 |
| 06/06/2005 | 1200 | 1200 | 10 06 07 | --52 47 50 |

## 3.2 Classification and diagnostics criteria for identifying SNRs

For optical spectral classification of SNRs we essentially followed the common scheme for identifying shock emission prescribed by Fesen, Blair & Kirshner (1995) (see also Blair & Long 1997) where the main attribute in optical spectra that makes a key difference between SNRs, H II regions and most PNe is if the ratio of [S II]/H$\alpha$ lines > 0.5. In practice the general optical spectral characteristics of all three groups of objects are broadly similar for certain emission lines and ratios (although physical processes can be very different). Consequently, there is now a recognised overlap at the extremes of object types in the various line ratio diagnostic diagrams that have been developed over the years, (e.g. Sabbadin et al. 1977 and Cantó 1981), to try to effectively discriminate between object types. Hence, a clear demarcation using such line ratio criteria alone is no longer always possible. A combination of morphological, environmental, photometric, spectroscopic and mutli-waveband data is now required in some difficult cases to resolve object identity ambiguity. Nevertheless, it is high ratios of the combined [S II] lines 6717 and 6731Å against H$\alpha$ that separates shock-heated gas, such as expected for an SNR from H II regions and the general PNe population. Some highly evolved PNe have now been found to exhibit high ratios of [S II]/H$\alpha$, in some cases exceeding unity (Pierce et al. 2004). Such ratios in PN emission spectra arise from shocks as the expanding PN shell collides with the ISM. Unusually high ratios of [N II]/H$\alpha$ of several are also seen in such cases in PNe where [S II]/H$\alpha$ is also enhanced (but never in H II regions) which may help to discriminate such instances from SNRs where such ratios are usually much lower (although this is not the case here). Fesen, Blair & Kirshner (1995) state that such basic emission line discriminators should also be supported with the presence of other typical SNR optical emission lines such as strong [O II] at 3727Å, [O I] at 6300 and 6364Å, [O III] at 4959 and 5007Å, [N II] lines at 6548 and 6584Å and the presence of Balmer lines. Most of these lines are also present in H II regions and PNe. The [O I] lines are uncommon in H II regions and photoionised nebulae such are PN, but their detection in SNR spectra is often compromised by the difficulty of properly subtracting these lines in the night sky emission spectrum where they are strong from the nebula spectra where they may be weak (see later). This is especially germane when the spectral resolution is low or moderate such that the kinematical SNR signatures are not sufficiently offset to aid subtraction.

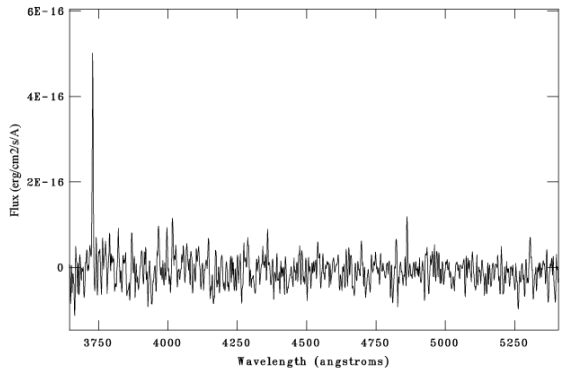 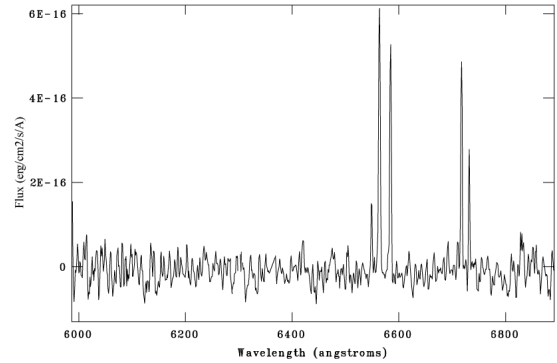

**Figure 4**: 1-D, flux calibrated blue spectrum for F1 slit position for G279.0+1.1 taken with the medium-resolution 600 lines/mm DBS grating. The strong [O II] line at 3727Å and weak H$\beta$ are clearly seen which strongly supports an SNR classification.

**Figure 5**: 1-D, flux calibrated red spectrum for F1 slit position for G279.0+1.1 taken with the higher-resolution 1200 lines/mm DBS grating. This red spectrum shows all the lines relevant for an SNR classification: [N II] at 6548 and 6584Å and strong [S II] at 6717 and 6731Å with the ratio [S II] / H$\alpha$ = 1.2 being typical for SNRs but not H II regions or PNe.

If we can confidently separate general SNR optical spectra from H II regions and in most cases PNe based on selected emission line ratios, there is still the question of discrimination against Wolf-Rayet (WR) star nebulae. Not only can these nebulae, consisting mainly of stellar ejecta from a central WR star, appear morphologically similar to the more coherent SNR ejecta in the form of arcs or rings but they may also exhibit very similar optical spectra due to the ejecta dynamics. A careful check against WR stars in the area in and around G279.0+1.1 in the 7$^{th}$ Catalogue of Galactic Wolf-Rayet stars (van der Hucht 2001) found no match with any of the newly detected filaments. Furthermore, the morphological structure and distribution of all 14 filaments and groups (which are, more or less, distributed in the direction from the centre of G279.0+1.1 towards the N-E limb, see Fig. 1) is different from typical WR star nebula. We therefore conclude that these optical nebulosities are most likely associated with G279.0+1.1.

### 3.2.1 Spectral properties of filament groups F1 and F3

Blue and red spectra taken at slit position F1 are shown in Figs. 4 and 5 and also presented in Table 3. An $F$ indicates the lines fluxes and resultant ratios are uncorrected for extinction while an $I$ indicates extinction correction has been applied. In the red F1 spectrum the ratio of [S II] /H$\alpha$ =1.2 classifies this filament well into the normal SNR domain (see also diagram by Cantó 1981). Additionally the [O I] 6300 and 6364Å lines are not clearly seen on 2D images. So, taking into account this fact and difficulty of achieving accurate sky-subtraction, these lines are not accepted for any analysis in this paper. The observed ratio of [N II] /H$\alpha$ = 1.1 is also well inside values seen in SNRs and PNe (but not H II regions). The [S II] 6717/6731 ratio is in the low density limit which is no more that can be expected for an old remnant like G279.0+1.1.

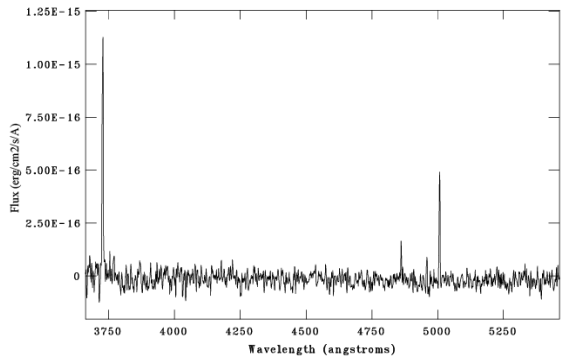 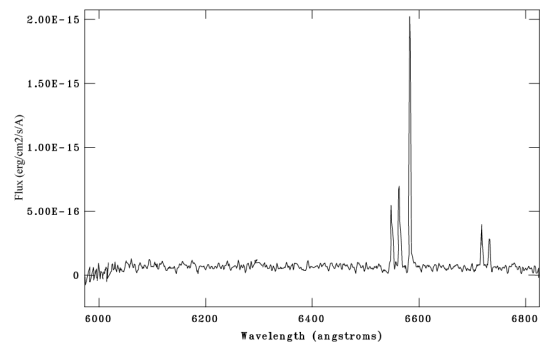

**Figure 6**: 1-D, flux calibrated blue spectrum for the F3 slit position on the N-E border of G279.0+1.1. Again, the medium-resolution 600 lines/mm grating was used. The extinction corrected [O II] line flux at 3727Å against H$\beta$ =100 gives an extremely high vale of 1270. Compared with the blue spectrum at slit position F1, where only H$\beta$ and [O II] at 3727Å are present, we now detect the two [O III] lines at 4959 and 5007Å which are common to SNRs.

**Figure 7**: 1-D, flux calibrated red spectrum for the F3 slit position. All the lines seen at the F1 slit position are also seen here but in different ratios: [S II] /H$\alpha$ = 0.7, [N II] /H$\alpha$ = 3.3. The [S II] 6717/6731 line ratio is ~1.5, again in the low density limit. It is the extremely high [N II] at 6584Å that make biggest distinction between these two red spectra. These line ratios rule out a H II region but not a highly evolved PNe (refer later discussion).

In the blue F1 spectrum shown in Fig. 4 only two lines are seen, very weak H$\beta$ and strong [O II] at 3727Å, where the ratio of this line against H$\beta$ = 100 gives a value of $I$ = 705 for extinction corrected and flux calibrated lines. Note the absence of [O III] effectively removes any possible confusion against an evolved PNe identification.

Application of these diagnostic line ratios to spectra taken at the F3 slit position on the rim of G279.0+1.1 also supports an SNR classification although there are some interesting differences in some of the line ratios compared with those from the F1 slit position. First, in the blue spectra, higher excitation [O III] lines are now detected and are relatively strong cf. H$\beta$ (see Fig. 6). Importantly, the key [O II] line at 3727Å is very strong and lends support to an SNR classification with $I$ = 1270 if H$\beta$ = 100.

**Table 3**: Ratio of the lines relative to H$\beta$ =100 for G279.0+1.1.

| $\lambda$ (Å) | Line | Flux (H$\beta$ =100) | | | |
|---|---|---|---|---|---|
| | | Slit pos. F1 [a] | | Slit pos. F3 [b] | |
| | | F | I | F | I |
| 3727 | [O II] | 473 | 705 | 802 | 1270 |
| 4861 | H$\beta$ | 100 | 100 | 100 | 100 |
| 4959 | [O III] | -- | -- | 84 | 81 |
| 5007 | [O III] | -- | -- | 308 | 289 |
| 6548 | [N II] | 137 | 83 | 382 | 215 |
| 6563 | H$\alpha$ | 476 | 286 | 515 | 286 |
| 6584 | [N II] | 385 | 227 | 1334 | 726 |
| 6717 | [S II] | 340 | 199 | 223 | 120 |
| 6731 | [S II] | 195 | 114 | 146 | 78 |
| Line Ratio | | | | | |
| 6717/6731 | | | LDL | | 1.5 |
| [N II] / H$\alpha$ | | | 1.1 | | 3.3 |
| [S II] / H$\alpha$ | | | 1.2 | | 0.7 |
| Extinction $c$ | | | 0.68 | | 0.79 |
| $E(B-V)$ | | | 0.47 | | 0.54 |
| | | F1 | | F3 | |
| log [H$\alpha$ /[S II]] | | -0.08 | | 0.15 | |
| log [H$\alpha$ /[N II]] | | -0.04 | | 0.52 | |
| Radial velocity [kms$^{-1}$] | | 28 $\pm$ 1.6 | | -40 $\pm$ 5.9 | |

[a] H$\beta$ flux= 4.9 $\times$ 10$^{-16}$ in units of erg cm$^{-2}$ s$^{-1}$ Å$^{-1}$
[b] H$\beta$ flux= 6 $\times$ 10$^{-16}$ in units of erg cm$^{-2}$ s$^{-1}$ Å$^{-1}$

In the red spectrum for the F3 slit position (Fig. 7) all the lines seen at the F1 slit position are also seen but in different ratios. The [N II]/H$\alpha$ ratio has a high value of ~3.3 while [SII]/H$\alpha$ = 0.7. These values are broadly consistent with an SNR identification (see Table 3) but inconsistent with H II regions which exhibit lower such ratios (Kennicutt et al 2000). However, as already noted, some highly evolved PNe can also have high [N II]/H$\alpha$ and even very occasionally [S II]/H$\alpha$ ratios approaching unity (e.g. Pierce et al. 2004). In PNe these ratios are seen in regions where the evolved PN shell is clearly interacting with the ISM. It should be noted that the F3 slit position is located at the NE edge of the remnant where an interaction with a denser ISM is probably taking place. The [S II] 6717/6731 of ~1.5 is again in the low density limit.

### 3.3 Compact radio source G278+0.8 and a new H II region

The known compact radio source G278.0+0.8 is within the overall boundary of G279.0+1.1 at the western extremity (see Fig. 1). Duncan et al. 1995 concluded that this strong source is most likely extragalactic and not related to G279.0+1.1 due to its small angular size (approximately 10 arcsec) and very steep radio spectrum ($\alpha$ = -0.9) estimated from the measured fluxes at 1.4, 2.4 and 4.85 GHz (see details in Duncan et al. 1995). G278.0+0.8 is not seen in H$\alpha$ light but we did find an excellent match of the diffuse radio cloud just to the west of G278.0+0.8 with diffuse H$\alpha$ emission in the same area with an approximate size of 20 arcmin (Fig. 8). However, rather than representing an optical detection of the western edge of the SNR we conclude, based on the pathological optical morphology, that it is most likely a new H II region (see right panel on Fig. 8) with bright emission rims seen against an obscured dust region. This dust is clearly seen in the IRAS data in Fig. 2 where a brighter and distinct 100 $\mu$m zone is co-incident. A check in SIMBAD[4] did not return any known optical nebular source in this area so we designate this H II region as SPF 0951-5310. Using the SHASSA flux calibrated H$\alpha$ survey of [et al. 2001] we obtained H$\alpha$ flux of this new proposed H II region as $\log F$ (H$\alpha$) = -9.41 erg cm$^{-2}$ s$^{-1}$, not corrected for emission of [N II] at 6548 and 6584Å (which is included in SHASSA bandpass) due to unknown flux of [N II] from the object. Note the newly discovered SNR filaments are unrecognised in the course of ~48 arcsec resolution of SHASSA survey.

SIMBAD did return a number of IRAS point sources and the ROSAT X-ray source 1RXS J095032.2-531402 (see Fig. 9) in the region of G278.0+0.8. We then checked this area in the IRAS Sky Survey.[5] First we checked across the general area of G279.0+1.1 and found that this SNR is not registered in any IRAS observing band. However, the region west of G278.0+0.8 is clearly seen in the infrared. This is shown in Fig. 9 where the IRAS morphological structure is very similar to the shell structure of an SNR. Unfortunately, due to the crude IRAS resolution of few arc minutes, it is very hard to confirm the nature of this shell structure. It is possible that we simply have IRAS point sources in a simple chain without any physical connection but apparently joined by the poor resolution. This structure, in the same or similar form, is seen not only at 100 $\mu$m, (Fig. 9), but also in the other IRAS observing bands of 12, 25 and 60 $\mu$m (see Fig. 2), although, as can be expected, there are brightness differences from source to source.

Further observations and analysis should confirm the real nature of the G278.0+0.8 emission cloud recognised in optical/radio/infrared and X-rays as well as any possible connection with SNR G279.0+1.1.

## 4 Discussion

Decent confirmatory optical spectra were obtained for several of the optical filaments that we believe are associated with the SNR G279.0+1.1. Given the slightly non photometric observing conditions, flux errors were estimated to be between 15% and 18% and with resulting extinction estimates $c$ at H$\beta$ of 0.68 and 0.79 and reddening $E(B-V)$ of 0.47 and 0.54 for slit positions F1 and F3 respectively (see Table 3).

---

[4] http://simbad.u-strasbg.fr/simbad/

[5] Actually, we used images of IRIS - Improved Reprocessing of the IRAS Survey. See http://irsa.ipac.caltech.edu/

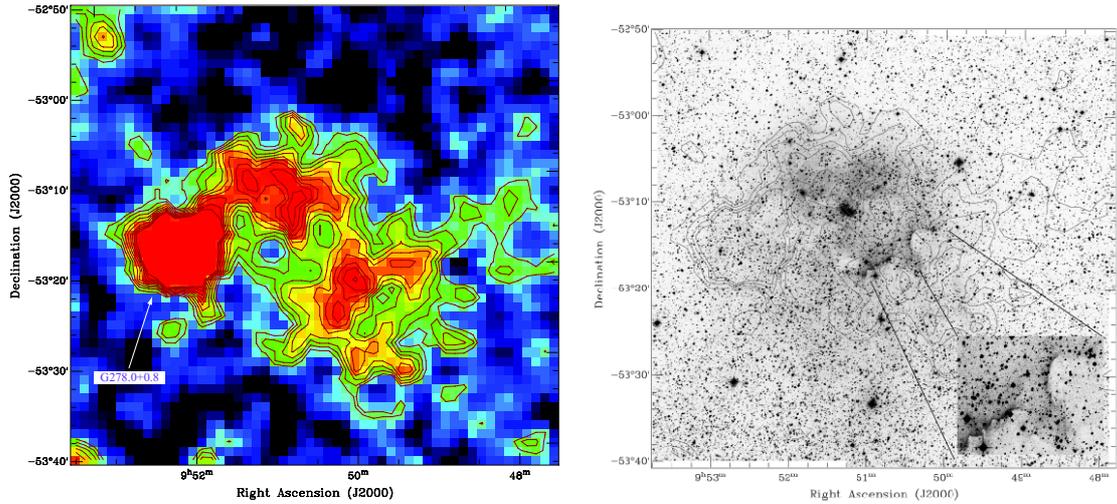

**Figure 8**: The left image is focused on G278.0+0.8 taken from the PMN survey (4.85 GHz) but including the area west from this compact object. Contours are chosen between 0.01 to 0.05 Jy beam$^{-1}$ (the signal is much stronger in the centre of G278.0+0.8) to clearly show the radio form of this emission cloud. The right panel shows the same area but in H$\alpha$ light showing an excellent match between the radio and optical images. Unfortunately, the compact radio object G278.0+0.8 is not seen in H$\alpha$ light. The inset image in the lower right corner of the H$\alpha$ image represents enlarged part of the strongest optical emission which is in the typical morphological form of a H II region with bright emission rims highlighted against dusty areas.

Due to the low level density observed for the F1 and F3 slit positions we were unable to measure a meaningful electron density assuming $T$ =10,000 K (Osterbrock & Ferland 2006). We can only conclude that the electron density for these two values are $>$10 cm$^3$, but could be as high as 470 cm$^3$ if we take into account errors in the measurement of the [S II] lines. For young SNRs the [S II] 6717 and 6731Å line ratio is usually less than 1, while for older remnants it is usually between 1 and the low density limit of ~1.5 as is the case here.

Stupar, Parker & Filipović (2007a) presented spectroscopic observations of evolved Galactic SNR G315.1+2.7 where unusually strong [N II] at 6584Å relative to H$\alpha$ was noticed at one but not the other of two slit positions across the same SNR filament separated by ~9 arcmin. Such ratios are highly unusual in evolved remnants so it was concluded that this most probably implies local enrichment and that the general Galactic nitrogen abundance gradient cannot apply. Support for this conclusion comes from F3 slit position from this work (Fig. 7) where a high ratio of [N II]/H$\alpha$ of 3.3 was observed. However the [N II]/H$\alpha$ ratio at the F1 slit position is inside the range typically exhibited by most optically detected SNR spectra.

The strong [N II] recorded at the F3 slit position can also be clearly seen in Fig. 10 where a 2-D spectral image of the same part of the spectrum as given in Fig. 7 is shown. The enhanced nitrogen seen across parts of evolved SNR G315.1+2.7 and now across parts of G279.0+1.1 are rare cases of localized nitrogen variability across an SNR due to interactions with the ISM. In the case of G279.0+1.1 two filaments with different [N II] at 6584Å are separated more than a degree. A few other SNRs with extremely strong [N II] lines at 6548 and 6584Å relative to H$\alpha$ have also been found in observed filaments such as: Kepler's SNR (Blair, Long & Vancura 1991), the Crab Nebula (MacAlpine et al. 1996), W50 (Kirshner & Chevalier 1980), G332.5-5.6 (or the 'Paperclip') (Stupar et al. 2007b), Pupis A (Dopita, Mathewson & Ford 1997) or in SNR candidate G247.8+4.9 (Zanin & Kerber 2000). From these cases it is clear that this apparent variation in nitrogen abundance, known to be present in young or moderate aged SNRs (Zanin & Kerber 2000) now been seen even in old remnants like G332.5-5.6 and now G279.0+1.1 which are most probably in the last, dissipation phase. G279.0+1.1 can therefore be classified amongst the small group of old Galactic remnants such as G332.5-5.6, Pupis A and G315.1+2.7 which have extremely strong [N II] at 6584Å. According to Kirshner & Chevalier (1980) the observed nitrogen strength has its origin in the SNRs interaction and sweeping up of the enriched material of the interstellar medium rather than in supernova ejecta itself.

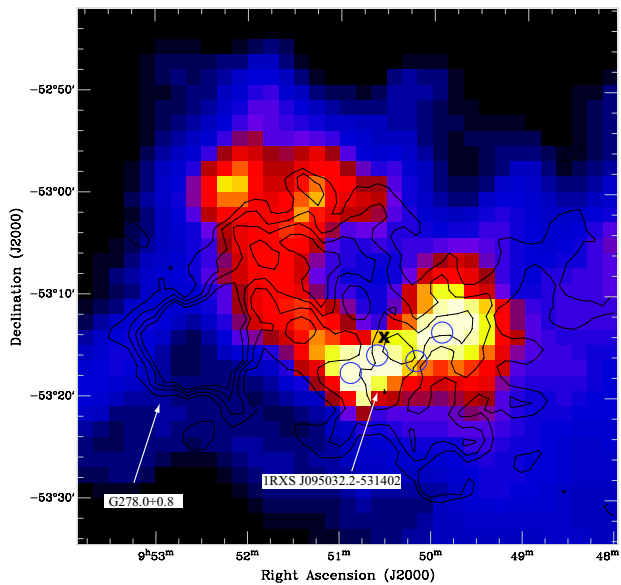

**Figure 9**: The area west of compact radio source G278.0+0.8 as in Fig. 8) but with the radio contours superimposed on the low resolution 100$\mu$m IRAS 'dust' image. An almost complete shell structure can be noticed in this infrared image which overlaps the extended 137 MJy/sr radio data. The IRAS point source IRAS 09490-5303 is found at the same location in Fig. 8 where, based on the SHS optical H$\alpha$ imaging, we suggest a new H II region is located. An excellent match can also be noticed between the IRAS source and ROSAT X-ray source 1RXS J095032.2-531402 (marked with an **X**) as well as a match with the radio and optical data from Fig. 8. The PMN contour levels are the same as in Fig. 8. Blue circles show the positions of the strongest IRAS point sources at 100$\mu$m: IRAS 09490-5303; IRAS 09488-5301; IRAS 09483-5302 and IRAS 09481-5259.

To obtain a velocity estimate of the shock we used the ratio of oxygen lines and H$\beta$ (Dopita et al. 1984)[6]. For the F1 slit position the [O II] line at 3727Å line was used and for F3, the [O III] line at 5007Å with resultant velocities of ~75 kms$^{-1}$ and ~90 kms$^{-1}$ respectively. This is in general agreement with Raymond (1979) where, if the [O III] 5007Å line is present, the shock velocity must be over 80 kms$^{-1}$. The observed ratio of [O III]/H$\beta$ ~4 for slit position F3 suggests that shocks with a complete recombination zone are present here and also support the estimated velocity for this area (see Raymond et al. 1988). Following this, we can estimate an upper value for the pre-shock cloud density for slit position F3 (using Dopita 1979) $\sim 13\, cm^3$.

---

[6]This is under pre-shock physical conditions of 10$^{-3}$ for density, temperature : 9,000K and 1 $\mu$G magnetic field strength.

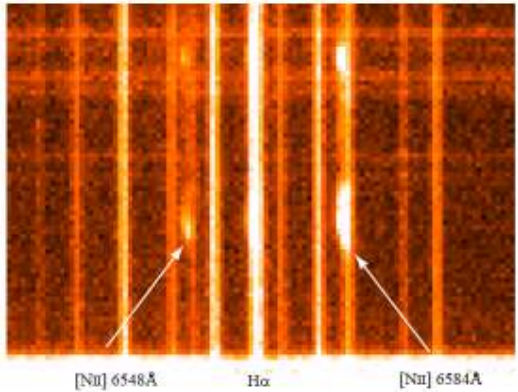

Figure 10: Part of 2-D (red) spectrum at slit position F3 (sky not subtracted) showing the strong but variable (localised) nebula blobs of [N II] at 6584Å which in ratio to H$\beta$ = 100 gives a value of $I(\lambda)$ = 726

We are unable to provide a kinematic distance estimate to this remnant as it is extremely difficult to derive the true spatial distribution of the radiative shock, usually seen optically in the form of filaments and often found within the radio boundaries of the SNR. Also we only have optical spectra sampling 2 of the 14 obviously associated filaments, even though individual filament velocities can be measured to a few km/s. Kinematic distance estimates can only be made in cases where many filaments are observed along the SNR rim and seen edge-on (viewed tangentially). Under such circumstances we can accept the measured radial velocities as representing the bulk motion of the remnant and measure the expansion velocity of the system. Of course this can be reasonably applied only when the SNR has the morphological form of a shell, which is not the case in our paper.

However, we have used the Mckee & Cowie (1975) technique to estimate the distance to G279.0+1.1, which connects blast wave energy and various cloud parameters. We assumed a canonical initial explosion energy of $10^{51}$ erg, a newly estimated diameter of 2.3°, an electron density of 10 cm$^3$ and the factor $\beta = 1$ (see Mckee & Cowie 1975) to derive a diameter of 120 pc and a distance of 3 kpc. Woermann & Jonas (1988), on the basis of their radio observations, and using $\Sigma - D$ system for distance estimates, concluded that the diameter of G279.0+1.1 is of the order of 100 pc at a distance of around 3 kpc, placing this remnant at the nearby Carina spiral arm and about 70 pc above the Galactic plane. Woerman & Jonas (1988) also analysed CO clouds in the area of this SNR which may be associated, and estimated a CO distance very close to that obtained by $\Sigma - D$ for the SNR. They also mentioned a [H I] feature overlapping the brighter part of G279.0+1.1 and estimated its kinematic distance as 8 kpc, but concluded that any association between G279.0+1.1 and the [H I] feature is unlikely. Duncan et al. (1995) also found three pulsars in the central area of this remnant, at distances between 2.5 and 4.9 kpc. Using transverse velocities and the estimated age of the pulsars they concluded that the pulsar marked as J1001-5507, at distance of $3.6 \pm 0.9$ kpc is the most likely candidate connected with G279.0+1.1, in good agreement with the other results.

Unfortunately, the above techniques of distance estimate are highly erratic: $\Sigma - D$ has at least a 40% error (e.g. see Stupar et al. 2005), while the Mckee & Cowie (1975) method depends on the initial explosion of progenitor star which, theoretically, can be between $E^{50}$ and $E^{51}$ erg, so these estimates, although all in reasonable agreement for distance and physical size, must still be treated with caution.

In summary, we have presented the first optically detected counterpart images to known SNR G279.0+1.1 together with confirmatory spectra consistent with shock excitation expected from such an evolved remnant. This is the first of many new optical detections of known remnants previously only detected in the radio now made possible with the SHS. A programme of confirmatory follow-up spectroscopy is underway and significant new optical data on known remnants is promised to help to further unravel their physical and kinematical characteristics.

## 5  Acknowledgements

We are grateful to the MSSSO Time Allocation Committees for enabling the spectroscopic follow-up to be obtained. MS is thankful to the Department of Physics, Macquarie University, Sydney, for travel support. We are thankfull to unknow referee for valuable comments.